\documentclass[prd,aps,showpacs,twocolumn,nofootinbib,floats,letterpaper,floatfix,groupedaddress,eqsecnum]{revtex4}

\usepackage{amsmath}
\usepackage{amssymb}
\usepackage{color}
\usepackage{graphicx}
\usepackage{epstopdf}         
\usepackage{latexsym}
\usepackage{times}
\usepackage{ifpdf}

\usepackage{dcolumn,epsfig}

\usepackage{amssymb,amsmath}

\def\be{\begin{equation}}
\def\ee{\end{equation}}
\def\beq{\begin{eqnarray}}
\def\eeq{\end{eqnarray}}

\def\IL{\relax{\rm I\kern-.18em L}}

\begin{document}

\title{Quasinormal modes of black holes in anti-de Sitter space: a numerical study of the eikonal limit}

\author{Jaqueline Morgan,$^{1} \footnote{Electronic address:
    jaqueline.morgan@ufabc.edu.br}$
     Vitor Cardoso,$^{2,3} \footnote{Electronic address:
    vitor.cardoso@ist.utl.pt}$
        Alex S. Miranda,$^{4} \footnote{Electronic address:
    astmiranda@if.ufrj.br}$
        C. Molina,$^{5} \footnote{Electronic address:
    cmolina@usp.br}$ and
        Vilson T. Zanchin$^{1} \footnote{Electronic address:
    zanchin@ufabc.edu.br}$}

\affiliation{${^1}$Centro de Ci\^encias Naturais e Humanas, Universidade Federal do ABC,
Rua Santa Ad\'elia 166, 09210-170 Santo Andr\'e, SP, Brazil}
\affiliation{${^2}$Centro Multidisciplinar de Astrof\'{\i}sica - CENTRA, Dept. de F\'{\i}sica,
Instituto Superior T\'ecnico, Av. Rovisco Pais 1, 1049-001 Lisboa, Portugal}
\affiliation{${^3}$Department of Physics and Astronomy, The University of Mississippi,
University, MS 38677-1848, USA}
\affiliation{${^4}$Instituto de F\'{i}sica, Universidade Federal do
Rio de Janeiro, Caixa Postal 68528, RJ 21941-972, Brazil}
\affiliation{${^5}$Escola de Artes, Ci\^{e}ncias e Humanidades, Universidade de
S\~{a}o Paulo, Avenida Arlindo Bettio 1000, 03828-000 S\~{a}o
Paulo, SP, Brazil}

\begin{abstract}
Using series solutions and time-domain evolutions, we probe the eikonal limit of the
gravitational and scalar-field quasinormal modes of large black holes and black branes
in anti-de Sitter backgrounds. These results are particularly relevant
for the AdS/CFT correspondence, since
the eikonal regime is characterized by the existence of long-lived modes which
(presumably) dominate the decay timescale of the perturbations.
We confirm all the main qualitative features of these slowly-damped modes as predicted by
Festuccia and Liu \cite{Festuccia:2008zx} for the scalar-field (tensor-type gravitational)
fluctuations. However, quantitatively we find dimensional-dependent correction factors.
We also investigate the dependence of the QNM frequencies on the horizon
radius of the black hole (brane) and the angular momentum (wavenumber) of
vector- and scalar-type gravitational perturbations.

\end{abstract}
\pacs{04.70.-s, 
11.25.Tq, 
11.10.Wx, 
04.50.-h, 
}

\maketitle

\section{Introduction}

The characteristic modes of black holes and black branes are eigenmodes
of these systems which convey important information about the background
geometry. These systems are intrinsically dissipative (energy flows to
the event horizon and/or towards the spatial infinity)
and therefore their eigenmodes are not stationary. One calls the eigenmodes
of these dissipative systems quasinormal modes (QNMs)
\cite{Berti:2009kk,Ferrari:2007dd,Kokkotas:1999bd,Nollert:1999ji}.
Within the anti-de Sitter/Conformal Field theory (AdS/CFT)
duality \cite{Maldacena:1997re,Witten:1998qj,
Gubser:1998bc}, these modes serve as an important tool for determining the
near-equilibrium properties of strongly coupled quantum field theories, in
particular their transport coefficients such as viscosity, conductivity and
diffusion constants \cite{Berti:2009kk,Son:2007vk}. Recently,
QNMs have also been used to study properties like the ``meson melting'' in
D3/D7-brane models \cite{Hoyos:2006gb,Myers:2007we,Myers:2008cj} and the spectrum of
collective excitations of holographic superconductors \cite{Amado:2009ts,Cubrovic:2009ye}.

It is thus of interest to understand the spectrum of black holes and black branes
in asymptotically anti-de Sitter (AdS) spacetimes. Recent studies by Festuccia and Liu
(henceforth FL) have shed light on this issue, predicting the existence of long-lived modes in asymptotically AdS
black hole (BH) geometries \cite{Festuccia:2008zx}. Their work shows that the eikonal limit in AdS depends sensitively on the relative size of the black hole. For small black holes, they find exponentially long-lived modes, which can be thought of as modes trapped inside the potential barrier.
If we write $\omega=\omega_R-i\omega_I$ for the typical ``energy''
eigenvalue, these modes take the Bohr-Sommerfeld form
\be
2i\int_{r_b}^{\infty}Q\,dr=\pi\left (2n+\frac{5}{2}\right)\,
,\quad n=0,\,1,\,...\label{bohrsommerfeld}
\ee
where
\be
Q\equiv \frac{1}{rf}\sqrt{(l+1/2)^2f-r^2\omega_R^2}\,,
\ee
$f$ is the black-hole horizon function (see Eq. (\ref{lineelementads})
below), and $l$ is the angular momentum of the perturbation. Accordingly,
their lifetime is dictated by a ``tunneling probability'' of the form
$\omega_I\propto \exp{(-2\int_{r_c}^{r_b}Q\,dr)}$, where $r_b$ and
$r_c<r_{b}$ are two real zeros (turning points) of $Q$, and the
proportionality coefficient is given in \cite{Berti:2009wx}. This
prediction is supported by numerical studies for small AdS black
holes \cite{Berti:2009wx}.

On the other hand, the nature of the long-lived modes for large black holes is completely different, since no trapped modes
are allowed in the large black-hole regime. A consequence of this fact is that
the ``Breit-Wigner resonance method'' used to investigate small AdS black holes
fails to give reliable results in the large black-hole regime \cite{Berti:2009wx}.
These modes are also expected to be long-lived \cite{Festuccia:2008zx} (see Section \ref{eikonal} below).
Thus these modes will presumably dominate the BH's response to arbitrary perturbations, hence
the thermalization timescale in the dual CFT. Since their existence may be
very relevant for the AdS/CFT conjecture, we decided to investigate
numerically these long-lived modes. We use two, well established methods
to study the QNMs of black holes. One is the series solution expansion
\cite{Horowitz:1999jd,Cardoso:2001bb,Cardoso:2003cj}, which outputs the characteristic
frequencies directly and is especially well suited for large black holes. By using
time-domain methods, in particular the scattering of Gaussian wavepackets \cite{Wang:2000dt,Wang:2004bv} (see also \cite{Chan:1999sc}),
we confirm and expand the series solution results and furthermore show that these
modes {\it are} excited in physically interesting situations. As a by-product, we confirm once more that there are no late-time tails in this geometry.

\section{\label{formal}Formulation of the problem}

\subsection{The background spacetime}

BH's in asymptotically AdS spacetimes form a class of solutions 
which is interesting from a theoretical point of view and central for the study of
strongly coupled field theories at finite temperature in the
gauge/gravity duality framework.
We are interested in a simple class of non-rotating, uncharged
$d$-dimensional Schwarzschild-AdS black holes with line element
\be ds^{2}= -fdt^{2}+f^{-1}dr^{2}+r^{2}d\Omega_{d-2}^2\,,
\label{lineelementads} \ee
where $f(r) = 1 + r^2/R^2 - r_0^{d-3}/r^{d-3}$ and $d\Omega_{d-2}^2$
is the metric of the unitary $(d-2)$-sphere. The AdS curvature radius
$R$ is related to the cosmological constant $\Lambda$ by
$R^2 = -(d-2)(d-1)/2\Lambda$. The
parameter $r_0$ is proportional to the mass $M$ of the
black hole:
$M=(d-2)A_{d-2}r_0^{d-3}/16\pi$,
where $A_{d-2} = 2 \pi^{(d-1)/2}/\Gamma{\left[(d-1)/2\right]}$. The
well-known Schwarzschild geometry corresponds to $R\to \infty$.
The black-hole horizon radius $r=r_+$ is the (unique) positive real root of $f(r)=0$.

In particular, we want to focus here on the large black hole regime,
$r_+/R \rightarrow \infty$. In this case, the above geometry
goes over to a $d-$dimensional plane-symmetric spacetime (black brane)
\cite{Horowitz:1999jd}, which is also an exact solution of
Einstein's equations \cite{Lemos:1994fn,Huang:1995zb,
Lemos:1994xp,Cai:1996eg}.
The black brane has the following line element
\be\label{fundo}
ds^2=-f(r)dt^{2}
+r^2\sum^{d-2}_{i=1}dx^{i}dx_{i}+\frac{1}{ f(r)}\;dr^{2}\,,
\ee
where $f(r)=r^{2}/R^{2}-r_{+}^{d-1}/(R^{2}r^{d-3})$. The
Hawking temperature of the black hole (\ref{lineelementads}) is
\be\label{hawkingTemp}
T=\frac{(d-1)r_{+}^2+(d-3)R^2}{4 \pi r_+ R^{2}}\,,
\ee
and it reduces to $T=(d-1)r_{+}/4 \pi R^{2}$ for large
black holes, which is the Hawking temperature of
the black brane (\ref{fundo}).

\subsection{Gravitational perturbations}
Gravitational perturbations in these backgrounds were considered in Ref.
\cite{Cardoso:2001bb,Miranda:2005qx,Miranda:2007bv} for $d=4$ and in
\cite{Kodama:2003jz,Ishibashi:2003ap} for higher dimensions. In a generic
number of dimensions, the gravitational perturbations can be divided
in three different types, the tensor-, vector- and scalar-type perturbations.
These can all be reduced to a master wave equation of the form
\be
f^2\frac{d^2\Psi}{dr^2}+ff'\frac{d\Psi}{dr}+
\left(\omega^2-V(r)\right )\Psi=0\,,\label{waveeq}
\ee
where the potential $V(r)$ depends on both the type of perturbation and
on basis functions used to separate the coordinates of the
($d-2$)-dimensional hypersurface of constant $r$ and $t$.
For instance, for tensor-type perturbations in the black-hole
background (\ref{lineelementads}),
\be
\frac{V}{f}=\frac{l(l+d-3)}{r^2}+\frac{(d-2)(d-4)}{4r^2}f+\frac{(d-2)f'}{2r}
\,.\label{potSAdS}
\ee
Here, the angular number $l$ is related to the eigenvalue of the hyper-spherical
functions used to factor out the dependence on the $(d-2)$-spherical angles.

For the black brane background (\ref{fundo}), the separation of trivial dimensions is achieved
through the Ansatz $e^{i\vec{q}.\vec{x}}$ and one ends up with the large
$r_+/R$ limit of (\ref{potSAdS}),
\be\label{pott}
\frac{V}{f}=\left[\frac{q^2}{r^2}+\frac{(d-2)(d-4)}{4r^2}f+\frac{(d-2)f'}{2r}\right].
\ee
The explicit form of $V(r)$ for the vector- and scalar-type perturbations can be found
in Refs. \cite{Cardoso:2001bb,Kodama:2003jz,Ishibashi:2003ap}.
The potential for tensor-type gravitational perturbations is equal to the potential
for scalar-field perturbations, so our results are also valid for spin-0 fields.

We will be interested in the large-$l,q$ limit of the QNM frequencies,
i.e., characteristic $\omega$'s for which the solutions to (\ref{waveeq}) satisfy
the appropriate boundary conditions. Note that in this limit one can formally
identify $q$ with $l$ in the black hole/black brane spacetimes. Hereafter we will
always refer to $q$, with the understanding that the replacement $q\to l$
describes large black holes.

An important characteristic of classical field evolutions on
asymptotically AdS spacetimes is the variety of choices for
the boundary conditions at spatial infinity. In general, these can
be Dirichlet, Neumann or Robin boundary conditions. So we need
to establish an objective criterion before to
choose a specific condition. In the AdS/CFT context,
a natural criterion is such that the QNM frequencies
of a certain field correspond to poles of two-point correlation
functions of the dual operator in the boundary field theory
\cite{Berti:2009kk,Nunez:2003eq,Kovtun:2005ev,Miranda:2008vb}.
 
When we consider a variable $\Psi$ such that the master
equation for the gravitational perturbations is of the
form (\ref{waveeq}), in general the Dirichlet condition at spatial infinity
is the `correct' boundary condition. There is only \textit{one}
exception: for scalar-type perturbations in four spacetime dimensions,
the boundary condition that leads to QNM frequencies
corresponding to poles of retarded correlation functions is of Robin
type (see Refs. \cite{Friess:2006kw,Michalogiorgakis:2006jc} for
a discussion in $d=4,5$ and \cite{Morgan:2009} for an
arbitrary number of dimensions.)

\subsection{\label{eikonal}Long-lived modes in the eikonal limit: analytical prediction}

In asymptotically AdS spacetimes the eikonal limit is especially interesting,
since large-$q$ modes can be very long-lived
\cite{Horowitz:1999jd,Festuccia:2008zx}. A WKB analysis suggests that for
the tensor-type gravitational perturbations (and therefore also scalar
fields) and $r_+/R\gg 1$ \cite{Festuccia:2008zx},
the following asymptotic behavior holds
\beq
R\,\omega^{\rm FL}&=&q+ \Pi_n \left (\frac{r_+}{R}\right )^
{\frac{2d-2}{d+1}}\,q^
{-\frac{d-3}{d+1}} \,,
\label{eikonalads}\\
\Pi_n &\equiv & 
\left (\sqrt{\pi}\left [\frac{d+1}{2}+2n\right ]\,
\frac{\Gamma\left(\frac{3d-1}{2d-2}\right)}
{\Gamma\left(\frac{1}{d-1}\right)}\right )
^{\frac{2d-2}{d+1}}e^{-\frac{2i\pi}{d+1}}\,,\nonumber
\eeq
as $q\rightarrow \infty$. So large-$q$ modes are very long-lived, and they
could play a prominent role in the BH's response to generic
perturbations. This is at variance with the asymptotically flat case, where
the damping timescale is roughly constant as $q$ varies. Notice also that the
scaling with the BH size differs from that of the weakly- and
highly-damped modes.

\section{\label{sec:numerics}Quasinormal frequencies}

\subsection{Methods}
%
\begin{figure}[ht]
\begin{tabular}{c}
\epsfig{file=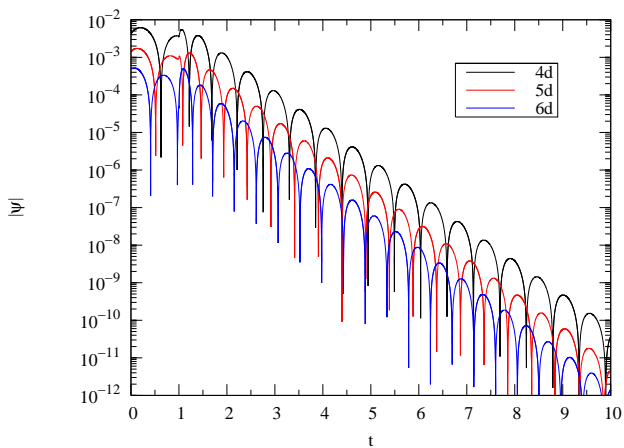,width=6cm,angle=270} 
\end{tabular}
\caption{Typical time-domain evolution of a gaussian wavepacket
resulting in ringdown (top to bottom $d=4,5,6$). Here we take $\psi (u,-30) = 0$,
$\psi (0,v) = \exp{\left[-(v -25)^2/18\right]}$ and $q=5$, where $u,v$ are standard null
coordinates \cite{Wang:2000dt,Wang:2004bv}. The signal is measured at
$r_{*}/R=-1$, where $dr/dr_{*}=f(r)$ and spatial infinity is at $r_*=0$.
\label{fig:timeevolution}}
\end{figure}
We use two conceptually different methods to determine the
gravitational QNM frequencies of the spacetimes
\eqref{lineelementads} and \eqref{fundo}, and both
methods yield consistent results, within the expected
error bars. The first consists on a series expansion method
\cite{Horowitz:1999jd,Berti:2009kk}, which reduces the problem
to finding roots of a polynomial. This method is
well suited to large black holes and black branes, though 
the convergence properties worsen for large wavenumber $q$.
In fact, for larger dimensions ($d>6$) and higher overtones
($n>1$ or $2$) the problems of convergence of the series solution
arise even for intermediate wavenumber values ($Rq/r_{+}\sim 10$).

The second method employed in this work consists on a direct
time-evolution in these backgrounds \cite{Wang:2000dt,Wang:2004bv}.
A particular example is shown in Figure \ref{fig:timeevolution},
which shows the time-development of a
Gaussian wavepacket in a $r_+/R=1$ black brane. The ringdown is
characterized by the decay timescale and ringing frequency, which can
directly be extracted from the slope and frequency of the signal
above. Equivalently, we characterize the QNMs by a
complex frequency $\omega=\omega_R-i\omega_I$.
Time evolution methods cannot determine very accurately which
overtone is dominating the response, though experience has shown
that the fundamental mode seems to be more excited than all
others, and by definition decays much slower \cite{Berti:2009kk}.
Most importantly, the scattering of wavepackets shows that
weakly-damped modes {\it are} excited in physically interesting situations.
Although not the direct focus of this work, our numerical results show no sign of a power-law tail at late stages,
confirming earlier predictions \cite{Horowitz:1999jd,Ching:1995tj}.

\subsection{Numerical Results}

\begin{figure}[ht]
\begin{tabular}{c}
\epsfig{file=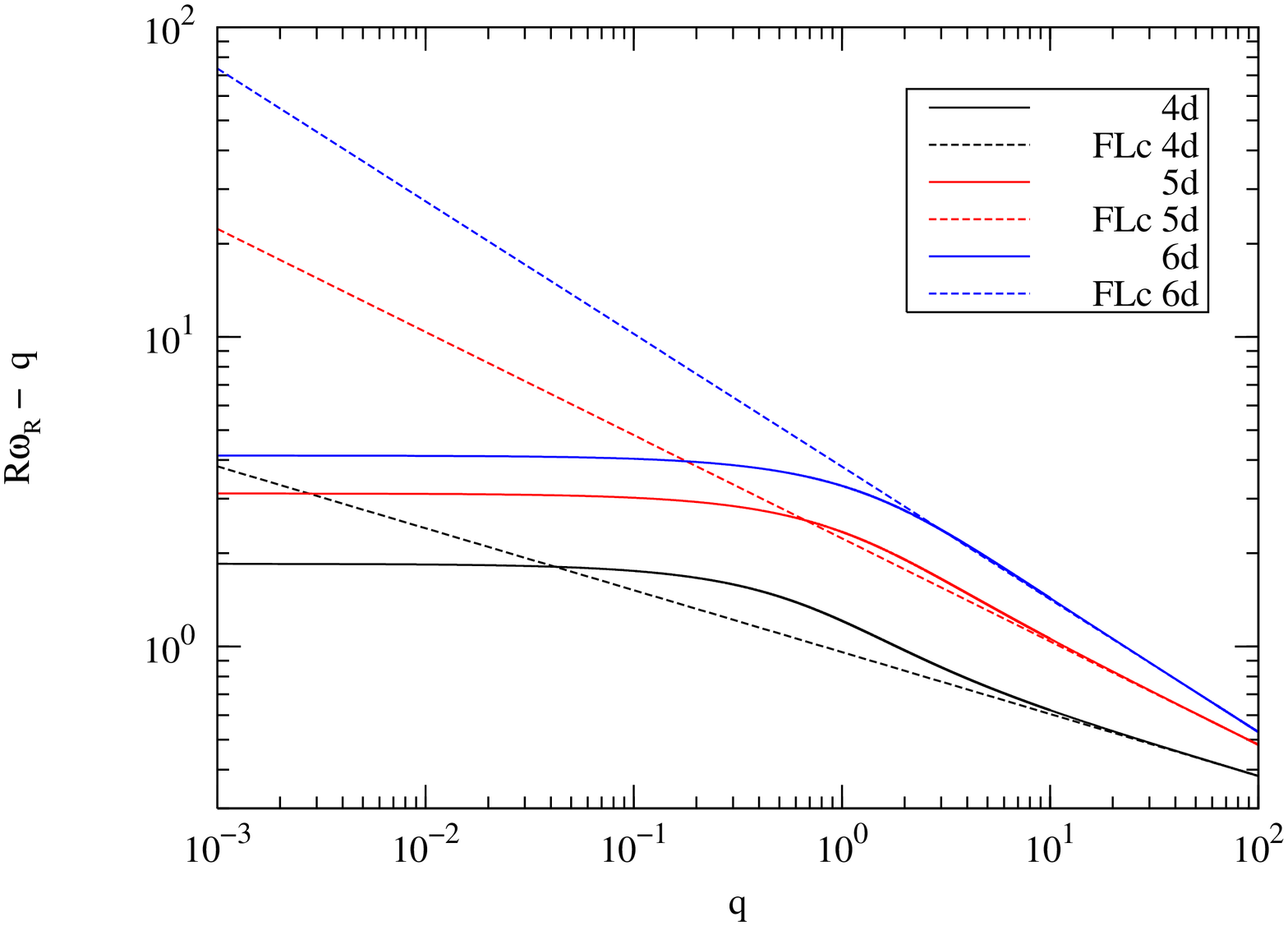,width=6cm,angle=270} \\ \epsfig{file=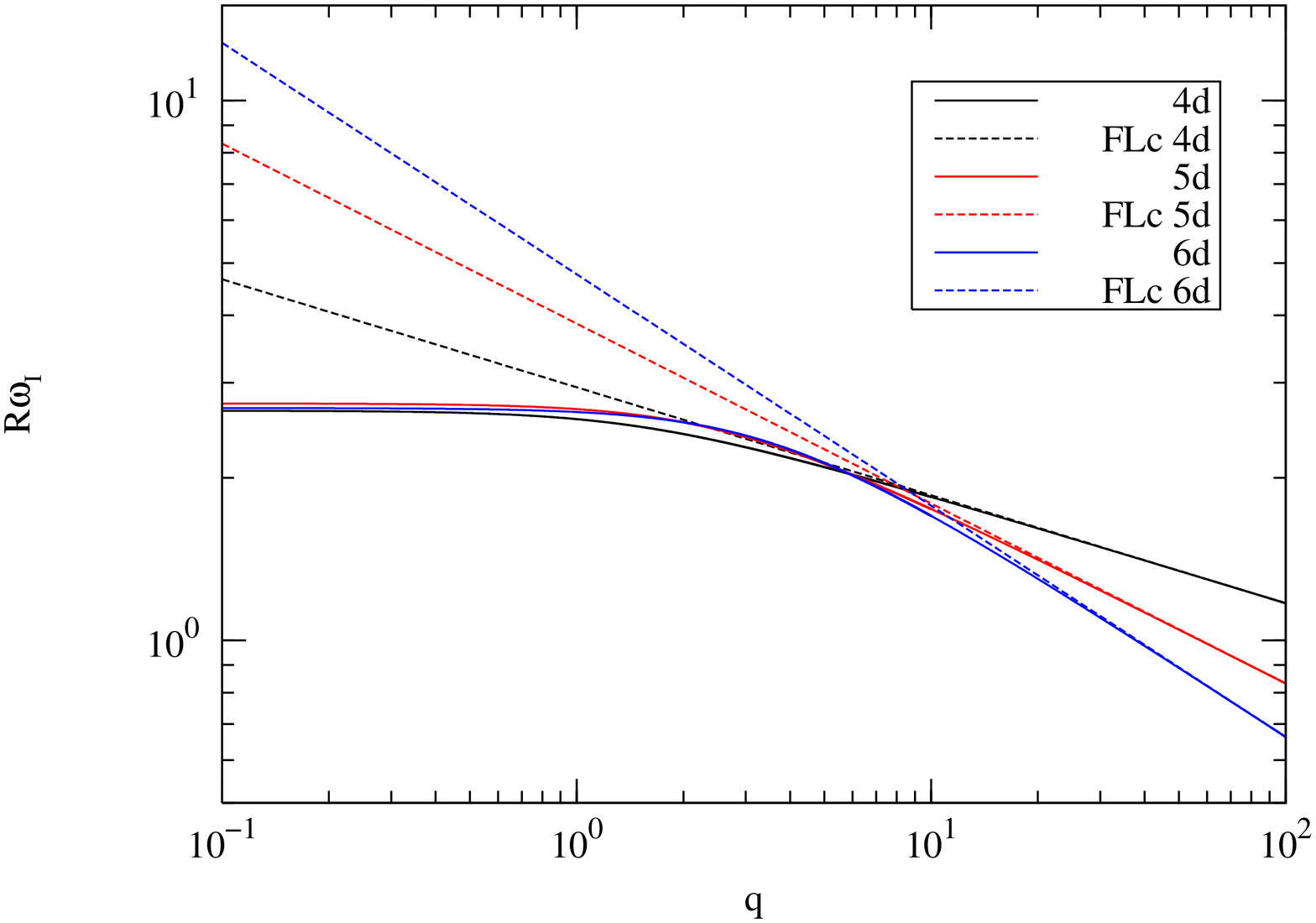,width=6cm,angle=270}
\end{tabular}
\caption{Numerical results for the fundamental scalar-field (tensor-type)
QNM frequencies of a $r_+/R=1$ black brane. Upper panel: real component $\omega_R$
(top to bottom are $d=6,5,4$).
Lower panel: imaginary component $\omega_I$ (top to bottom for $q>10$ corresponds to $d=4,5,6$).
Dotted lines are the analytical prediction (\ref{eikonalads}),
corrected by a prefactor $a$ shown in Table \ref{tab:summary}.
\label{fig:SAdSs}}
\end{figure}
In Figure \ref{fig:SAdSs} we show numerical results for
scalar-field (tensor-type gravitational) perturbations of
a $r_+/R=1$ black brane. Similar results hold for large black holes. 
Low-$q$ results for $d=4,5$ are well known in the
literature (See Ref. \cite{Berti:2009kk} and references therein),
while the higher-dimensional cases are discussed in details
in Ref. \cite{Morgan:2009}:
both $\omega_R$ and $\omega_I$ are almost independent on $q$ (or $l$ for
large black holes) in this regime. For wavenumbers $q \gg r_+/R$, the
qualitative behavior changes: $\omega_R$ grows linearly while $\omega_I$
decreases with $q$. Furthermore, it is also clear from Fig. \ref{fig:SAdSs}
that both the sub-leading term in $\omega_R$ and the leading term in $\omega_I$
scale as a power of $q$ (or $l$). This power can be directly read from the slope of
the curves of Figure \ref{fig:SAdSs} at large $q$. To investigate this further,
we parameterize the numerical results by
\be
R\,\omega_{R}=q+\alpha_{R}\,q^{-\beta_{R}}\,,
\qquad R\,\omega_{I}=\alpha_{I}\,q^{-\beta_{I}}\,,
\ee
and we extract $\beta_{R,I}$ by least-squares. We obtain the values
listed in Table \ref{tab:exponent}, where we also show the prediction
by FL, i.e., $\beta_{R,I}=(d-3)/(d+1)$. 
\begin{table}[h]
\caption{\label{tab:exponent}Best fit to exponents $\beta_{R,I}$, and FL's
prediction, $\frac{d-3}{d+1}$.}
\begin{tabular}{ccccccccccc}
\hline 
Type & $n$ $\;\;$
&\multicolumn{3}{c}{$4d$}&\multicolumn{3}{c}{$5d$}
&\multicolumn{3}{c}{$6d$}\\
          & &$\beta_R$&$\beta_I$&FL$\;\;\;$&
$\beta_R$&$\beta_I$&FL$\;\;\;$& $\beta_R$&$\beta_I$&FL\\ \hline
\hline\\
         
&0$\;\;$&0.20&0.20&0.20$\;\;\;$&0.33&0.33&0.33$\;\;\;$&0.43&0.43
&0.43
\\
         
&1$\;\;$&0.21&0.20&0.20$\;\;\;$&0.34&0.33&0.33$\;\;\;$&0.43&0.43
&0.43\\
 tensor  
&2$\;\;$&0.21&0.20&0.20$\;\;\;$&0.34&0.33&0.33$\;\;\;$&0.43&0.43
&0.43\\
         
&3$\;\;$&0.22&0.19&0.20$\;\;\;$&0.34&0.33&0.33$\;\;\;$&0.43&0.42
&0.43\\
         
&4$\;\;$&0.24&0.20&0.20$\;\;\;$&0.34&0.32&0.33$\;\;\;$&0.43&0.41
&0.43\\\\\hline\\

&0$\;\;$&0.19&0.20&0.20$\;\;\;$&0.33&0.34&0.33$\;\;\;$&0.42&0.44
&0.43\\
vector   
&1$\;\;$&0.20&0.20&0.20$\;\;\;$&0.33&0.33&0.33$\;\;\;$&0.43&0.43
&0.43\\
         
&2$\;\;$&0.20&0.20&0.20$\;\;\;$&--&--&0.33$\;\;\;$&--&--&0.43
\\\\\hline\\
          & 0$\;\;$ & 0.17&0.21 & 0.20 $\;\;\;$  & 0.30&0.31 & 0.33
$\;\;\;$& 0.39  & 0.41&0.43 \\
scalar    & 1$\;\;$ & 0.16&0.22 & 0.20 $\;\;\;$  & 0.31&0.35 & 0.33
$\;\;\;$ & --  & --&0.43 \\
          & 2$\;\;$ & 0.19&0.21 & 0.20 $\;\;\;$& 0.31&0.34 & 0.33
$\;\;\;$ & --  &
--&0.43 \\\hline
\end{tabular}
\end{table}

Our numerical results are consistent with a
$q^{-(d-3)/(d+1)}$ dependence of the characteristic frequencies,
not only for the dimensions shown in Table \ref{tab:exponent}, but
also for $d=7$, $8$ and $9$. Furthermore, we computed the same modes
for a $r_+/R=100$ black hole, and to numerical accuracy
we get the same results after a
rescaling by $\left(\frac{r_+}{R}\right )^{\frac{2d-2}{d+1}}$ is
performed. Thus, our results are also highly consistent with the
functional dependence on $r_+,q$ as given by equation
(\ref{eikonalads}). This agreement is nicely illustrated in Fig.
\ref{fig:SAdSs}. In a log-log plot, the analytical result predicts a
line with slope $-\frac{d-3}{d+1}$, which overlaps very well with the
numerical results for large $q$.

We now {\it assume} the power-law behavior (\ref{eikonalads}) in
$q$ and $r_+$, and fit the numerical results to the following function
\be
R\,\omega^{\rm Num}=1+ \left(a_R\,{\rm Re}[\Pi_n]+i a_I\,{\rm Im}[\Pi_n]
\right)\left (\frac{r_+}{R}\right )^{\frac{2d-2}{d+1}}\,q^{-\frac{d-3}{d+1}} \,,
\ee
thereby testing the prefactor in (\ref{eikonalads}). If
Eq.(\ref{eikonalads}) captures correctly all of the features of these
modes, then $a_R\approx a_I \approx 1$. Table \ref{tab:summary} summarizes our
main results, with the correction factors to (\ref{eikonalads}) for
each dimension number $d$ and gravitational sector. The results
in Table \ref{tab:summary} are strong indicators that
Eq.(\ref{eikonalads}) does {\it not} account for the correct
quantitative behavior of these weakly-damped modes. 
\begin{table}[h]
\caption{\label{tab:summary}Correction factors to the analytical formula
\eqref{eikonalads}.}
\begin{tabular}{cccccccc}
\hline 
Type & $n$ $\quad$ &\multicolumn{2}{c}{$4d$}&
\multicolumn{2}{c}{$5d$} &\multicolumn{2}{c}{$6d$}\\
          & & $a_R$ & $a_I$ $\quad$& $a_R$ & $a_I$ & $a_R$ & $a_I$\\
\hline \hline\\
          & 0$\quad$ & 1.83 & 1.82 $\quad$& 3.00 & 2.99$\quad$ & 4.56
&4.55 \\
          & 1$\quad$ & 1.86 & 1.85 $\quad$& 3.12 & 3.10 $\quad$& 4.82
&4.80\\
tensor& 2$\quad$ & 1.87 & 1.86$\quad$ & 3.15 & 3.13$\quad$ & 4.86
&4.90\\
          & 3$\quad$ & 1.88 & 1.86$\quad$ & 3.17 & 3.14$\quad$ & 4.94
&4.89\\
          
          & 4$\quad$ & 1.86 & 1.85$\quad$ & 3.18 & 3.14$\quad$ & 4.96
&4.89\\\hline\\
          & 0$\quad$ & 1.00 & 1.02 $\quad$& 1.80 & 1.82$\quad$ & 2.92
&2.95\\
vector  & 1$\quad$ & 1.38 & 1.38 $\quad$& 2.34 & 2.35$\quad$& 3.68
&3.69\\
          & 2$\quad$ & 1.53 & 1.53$\quad$ & - & -$\quad$ & - & -
\\\hline\\
          & 0$\quad$ & 0.29 & 0.30 $\quad$& 0.77 & 0.80 $\quad$& 1.59
& 1.64 \\
scalar    & 1$\quad$ & 0.92 & 0.93$\quad$ & 1.59 & 1.65 $\quad$& - & -
\\
          & 2$\quad$ & 1.20 & 1.20 $\quad$& 2.02 & 2.04$\quad$ & - & -
\\\hline
\end{tabular}
\end{table}
The results are consistent with {\it real}, overtone-independent,
but dimension-dependent correction factors for scalar-field
(tensor-type gravitational) perturbations. This correction factor
grows with $d$ and might become dominant at large $d$.

Equation (\ref{eikonalads}) is not supposed to hold for vector-type
and scalar-type gravitational perturbations, but we find it captures
the essential qualitative behavior with $r_+$, $q$. It can describe
quantitatively the numerical results if multiplied by a real constant,
which depends on the overtone $n$ and the spacetime dimension $d$.
This clearly suggests a new form for $\Pi_n$.

\section{Conclusions and outlook}

Our numerical results lend strong support to FL's prediction for
the existence of long-lived modes in the eikonal limit. Furthermore, the
functional dependence of these modes on the horizon radius and momentum $q$ is
consistent with FL, but we show that the correct quantitative
behavior is not. In particular, if we correct their prediction by
some (real) correction factors, listed in Table \ref{tab:summary},
one can account extremely well for the numerical results. Taken together,
our results suggest that large-$q$ tensor-type (or scalar-field)
quasinormal frequencies of black holes and black branes are described by
\be
R\,\omega=q+ a\,\Pi_n \left (\frac{r_+}{R}\right )^
{\frac{2d-2}{d+1}}\,q^{-\frac{d-3}{d+1}} \,,\label{finalguido}
\ee
where the value of $a$ depends on $d$, but it is independent on
the overtone number $n$.

We also provide correction factors that would make prediction
(\ref{eikonalads}) describe well other types of gravitational
quasinormal modes (vector and scalar). For such perturbations,
the real constant $a$ depends not only on $d$ but also on $n$,
suggesting a completely different form to $\Pi_n$. In any case, 
there is a simple and universal dependence on $r_+$ and $q,l$ in this eikonal
regime. Perhaps a simple interpretation in terms of geodesics can be given,
as is done in asymptotically flat spacetimes \cite{Berti:2009kk,Cardoso:2008bp}.
Clearly, more analytical and numerical studies are necessary to have a clear
picture of the eikonal, weakly-damped regime of quasinormal modes
of large black holes and black branes. A particularly interesting direction is to assess the degree to which these modes can be excited,
which is tantamount to computing the residue of the Green function at the QNM pole.
This is an important research topic in asymptotically flat spacetime, where it allows to predict how astrophysical black holes respond to external sources \cite{Berti:2006wq}, and has also recently started to be explored in the gauge/gravity duality scenario \cite{Amado:2008ji,Amado:2007yr}.

\noindent {\bf Note added in proof:} We have recently been informed \cite{guido} of a mistake in one
integral in FL, which introduces the correction factor $a=(1/2)(d-1)^{(2d-2)/(d+1)}$ in Eq. (\ref{finalguido}).
This correction factor is consistent with all our numerical results for scalar fields or tensor-type gravitational perturbations.

\section*{Acknowledgements}
We would like to thank Guido Festuccia for helpful correspondence.
This work is partially supported by Funda\c c\~ao para a Ci\^encia e Tecnologia (FCT)
- Portugal through project PTDC/FIS/64175/2006 and by Conselho Nacional de
Desenvolvimento Cient\'\i fico e Tecnol\'ogico of Brazil
(CNPq). JM thanks Funda\c c\~ao Universidade Federal do ABC (UFABC) for a grant.



\begin{thebibliography}{99}

\bibitem{Berti:2009kk}
  E.~Berti, V.~Cardoso and A.~O.~Starinets,
  arXiv:0905.2975 [gr-qc].

\bibitem{Ferrari:2007dd}
  V.~Ferrari and L.~Gualtieri,
  Gen.\ Rel.\ Grav.\  {\bf 40}, 945 (2008)
  [arXiv:0709.0657 [gr-qc]].

\bibitem{Kokkotas:1999bd}
  K.~D.~Kokkotas and B.~G.~Schmidt,
  Living Rev.\ Rel.\  {\bf 2}, 2 (1999)
  [arXiv:gr-qc/9909058].

\bibitem{Nollert:1999ji}
  H.~P.~Nollert,
  Class.\ Quant.\ Grav.\  {\bf 16}, R159 (1999).

\bibitem{Maldacena:1997re}
  J.~M.~Maldacena,
  Adv.\ Theor.\ Math.\ Phys.\  {\bf 2}, 231 (1998)
  [Int.\ J.\ Theor.\ Phys.\  {\bf 38}, 1113 (1999)]
  [arXiv:hep-th/9711200].

\bibitem{Witten:1998qj}
  E.~Witten,
  Adv.\ Theor.\ Math.\ Phys.\  {\bf 2}, 253 (1998)
  [arXiv:hep-th/9802150].

\bibitem{Gubser:1998bc}
  S.~S.~Gubser, I.~R.~Klebanov and A.~M.~Polyakov,
  Phys.\ Lett.\  B {\bf 428}, 105 (1998)
  [arXiv:hep-th/9802109].

\bibitem{Son:2007vk}
  D.~T.~Son and A.~O.~Starinets,
  Ann.\ Rev.\ Nucl.\ Part.\ Sci.\  {\bf 57}, 95 (2007)
  [arXiv:0704.0240 [hep-th]].

\bibitem{Hoyos:2006gb}
  C.~Hoyos-Badajoz, K.~Landsteiner and S.~Montero,
  JHEP {\bf 0704}, 031 (2007)
  [arXiv:hep-th/0612169].

\bibitem{Myers:2007we}
  R.~C.~Myers, A.~O.~Starinets and R.~M.~Thomson,
  JHEP {\bf 0711}, 091 (2007)
  [arXiv:0706.0162 [hep-th]].

\bibitem{Myers:2008cj}
  R.~C.~Myers and A.~Sinha,
  JHEP {\bf 0806}, 052 (2008)
  [arXiv:0804.2168 [hep-th]].

\bibitem{Amado:2009ts}
  I.~Amado, M.~Kaminski and K.~Landsteiner,
  JHEP {\bf 0905}, 021 (2009)
  [arXiv:0903.2209 [hep-th]].

\bibitem{Cubrovic:2009ye}
  M.~Cubrovic, J.~Zaanen and K.~Schalm,
  arXiv:0904.1993 [hep-th].

\bibitem{Festuccia:2008zx}
  G.~Festuccia and H.~Liu,
  arXiv:0811.1033 [gr-qc].

\bibitem{Berti:2009wx}
  E.~Berti, V.~Cardoso and P.~Pani,
  arXiv:0903.5311 [gr-qc].

  
\bibitem{Horowitz:1999jd}
  G.~T.~Horowitz and V.~E.~Hubeny,
  Phys.\ Rev.\  D {\bf 62}, 024027 (2000)
  [arXiv:hep-th/9909056].

\bibitem{Cardoso:2001bb}
  V.~Cardoso and J.~P.~S.~Lemos,
  Phys.\ Rev.\  D {\bf 64}, 084017 (2001)
  [arXiv:gr-qc/0105103].
  
\bibitem{Cardoso:2003cj}
  V.~Cardoso, R.~Konoplya and J.~P.~S.~Lemos,
  Phys.\ Rev.\  D {\bf 68}, 044024 (2003)
  [arXiv:gr-qc/0305037].

\bibitem{Wang:2000dt}
  B.~Wang, C.~Molina and E.~Abdalla,
  Phys.\ Rev.\  D {\bf 63}, 084001 (2001)
  [arXiv:hep-th/0005143].
  
\bibitem{Wang:2004bv}
  B.~Wang, C.~Y.~Lin and C.~Molina,
  Phys.\ Rev.\  D {\bf 70}, 064025 (2004)
  [arXiv:hep-th/0407024].

\bibitem{Chan:1999sc}
  J.~S.~F.~Chan and R.~B.~Mann,
  Phys.\ Rev.\  D {\bf 59}, 064025 (1999).
  
\bibitem{Lemos:1994fn}
  J.~P.~S.~Lemos,
  Class.\ Quant.\ Grav.\  {\bf 12}, 1081 (1995)
  [arXiv:gr-qc/9407024].

\bibitem{Huang:1995zb}
  C.~G.~Huang and C.~B.~Liang,
  Phys.\ Lett.\  A {\bf 201}, 27 (1995).

\bibitem{Lemos:1994xp}
  J.~P.~S.~Lemos,
  Phys.\ Lett.\  B {\bf 353}, 46 (1995)
  [arXiv:gr-qc/9404041].

\bibitem{Cai:1996eg}
  R.~G.~Cai and Y.~Z.~Zhang,
  Phys.\ Rev.\  D {\bf 54}, 4891 (1996)
  [arXiv:gr-qc/9609065].

\bibitem{Miranda:2005qx}
  A.~S.~Miranda and V.~T.~Zanchin,
  Phys.\ Rev.\  D {\bf 73}, 064034 (2006)
  [arXiv:gr-qc/0510066].

\bibitem{Miranda:2007bv}
  A.~S.~Miranda and V.~T.~Zanchin,
  Int.\ J.\ Mod.\ Phys.\  D {\bf 16}, 421 (2007).

\bibitem{Kodama:2003jz}
  H.~Kodama and A.~Ishibashi,
  Prog.\ Theor.\ Phys.\  {\bf 110}, 701 (2003)
  [arXiv:hep-th/0305147].

\bibitem{Ishibashi:2003ap}
  A.~Ishibashi and H.~Kodama,
  Prog.\ Theor.\ Phys.\  {\bf 110}, 901 (2003)
  [arXiv:hep-th/0305185].

\bibitem{Nunez:2003eq}
  A.~Nunez and A.~O.~Starinets,
  Phys.\ Rev.\  D {\bf 67}, 124013 (2003)
  [arXiv:hep-th/0302026].

\bibitem{Kovtun:2005ev}
  P.~K.~Kovtun and A.~O.~Starinets,
  Phys.\ Rev.\  D {\bf 72}, 086009 (2005)
  [arXiv:hep-th/0506184].


\bibitem{Miranda:2008vb}
  A.~S.~Miranda, J.~Morgan and V.~T.~Zanchin,
  JHEP {\bf 0811}, 030 (2008)
  [arXiv:0809.0297 [hep-th]].

\bibitem{Friess:2006kw}
  J.~J.~Friess, S.~S.~Gubser, G.~Michalogiorgakis and S.~S.~Pufu,
  JHEP {\bf 0704}, 080 (2007)
  [arXiv:hep-th/0611005].

\bibitem{Michalogiorgakis:2006jc}
  G.~Michalogiorgakis and S.~S.~Pufu,
  JHEP {\bf 0702}, 023 (2007)
  [arXiv:hep-th/0612065].

\bibitem{Ching:1995tj}
  E.~S.~C.~Ching, P.~T.~Leung, W.~M.~Suen and K.~Young,
  Phys.\ Rev.\  D {\bf 52}, 2118 (1995)
  [arXiv:gr-qc/9507035].

\bibitem{Morgan:2009}
  J.~Morgan, V.~Cardoso, A.~S.~Miranda, C.~Molina and V.~T.~Zanchin,
  In preparation.

\bibitem{Cardoso:2008bp}
  V.~Cardoso, A.~S.~Miranda, E.~Berti, H.~Witek and V.~T.~Zanchin,
  Phys.\ Rev.\  D {\bf 79}, 064016 (2009)
  [arXiv:0812.1806 [hep-th]].
  
\bibitem{Berti:2006wq}
  E.~Berti and V.~Cardoso,
  Phys.\ Rev.\  D {\bf 74}, 104020 (2006)
  [arXiv:gr-qc/0605118].

\bibitem{Amado:2008ji}
  I.~Amado, C.~Hoyos-Badajoz, K.~Landsteiner and S.~Montero,
  JHEP {\bf 07}, 133 (2008)
  [arXiv:0805.2570 [hep-th]].

\bibitem{Amado:2007yr}
  I.~Amado, C.~Hoyos-Badajoz, K.~Landsteiner and S.~Montero,
  Phys.\ Rev.\  D {\bf 77}, 065004 (2008)
  [arXiv:0710.4458 [hep-th]].

\bibitem{guido} G. Festuccia, private communication.

\end{thebibliography}
\end{document}